\documentclass[11pt]{article}
\usepackage{graphicx}
\usepackage{amsmath,amssymb}

\newcommand{\be}{\begin{equation}}
\newcommand{\ee}{\end{equation}}
\newcommand{\bea}{\begin{eqnarray}}
\newcommand{\eea}{\end{eqnarray}}
\newcommand{\fatk}{\mathbf k}

\newcommand{\fate}{\mathbf e}
\newcommand{\mat}{\left ( \begin{array}{cc}}
\newcommand{\emat}{\end{array} \right )}

\newcommand{\bfe}{{\mathbf e}}

\newcommand{\bfk}{{\mathbf k}}
\newcommand{\bfl}{{\mathbf l}}

\newcommand{\bfq}{{\mathbf q}}
\newcommand{\bfr}{{\mathbf r}}

\newcommand{\bfmu}{{\boldsymbol \mu}}
\newcommand{\bfnu}{{\boldsymbol \nu}}

\newcommand{\bD}{\bar{D}}

\newcommand{\bphi}{{\bar{\phi}}}
\newcommand{\bpsi}{\bar{\psi}}

\newcommand{\bxi}{\bar{\xi}}

\newcommand{\bareta}{\bar{\eta}}

\newcommand{\hmu}{\hat{\mu}}

\newcommand{\cA}{{\cal A}}
\newcommand{\cD}{{\cal D}}

\newcommand{\cL}{{\cal L}}
\newcommand{\cN}{{\cal N}}

\newcommand{\cU}{{\cal U}} 

\newcommand{\bcU}{\overline{\cal U}}
\newcommand{\bcD}{\overline{\cal D}}

\newcommand{\Z}{\mathbb{Z}}

\newcommand{\nn}{\nonumber}

\newcommand{\Tr}{{\rm Tr}\,}

\newcommand{\del}{\partial}
\newcommand{\e}{\epsilon}

\begin{document}
\begin{titlepage}

\setcounter{page}{0}
\renewcommand{\thefootnote}{\fnsymbol{footnote}}

\vspace{20mm}

\begin{center}
{\large\bf
Geometry of Orbifolded Supersymmetric Lattice Gauge Theories
}

\vspace{20mm}
Poul H. Damgaard%
\footnote{\tt phdamg@nbi.dk}  
and So Matsuura%
\footnote{\tt matsuura@nbi.dk}\\
\vspace{10mm}

{\em The Niels Bohr Institute, \\
The Niels Bohr International Academy,\\
Blegdamsvej 17, DK-2100 Copenhagen, Denmark}

\end{center}

\vspace{20mm}
\centerline{{\bf Abstract}}
\vspace{10mm}

We prove that the prescription for construction of
supersymmetric lattice gauge theories by orbifolding
and deconstruction directly leads to
Catterall's geometrical discretization scheme in general.
These two prescriptions always give the same lattice 
discretizations when applied to theories of $p$-form fields. 
We also show that the geometrical discretization scheme 
can be applied to more general theories.

\end{titlepage}
\newpage

\renewcommand{\thefootnote}{\arabic{footnote}}
\setcounter{footnote}{0}

Among the many recent developments towards 
putting exactly preserved supersymmetries on a 
space-time lattice, 
one of the most striking results is that apparently
quite different formulations are related 
to each other. It appears that the orbifolding procedure
is the unifying framework  \cite{Cohen:2003xe}%
--\nocite{Cohen:2003qw}\cite{Kaplan:2005ta}. 
{}For example, it has been shown in \cite{Damgaard:2007xi} 
that Catterall's complexified lattice theories 
\cite{Catterall:2003wd}--%
\nocite{Catterall:2004np}\cite{Catterall:2005fd}
constructed by a
geometrical discretization scheme from continuum theories
in the twisted formulation 
can be reproduced using the orbifolding procedure\footnote{ 
{}For further analysis, 
see, $e.g.$, refs. \cite{Catterall:2006jw}%
--\nocite{Giedt:2003ve}\nocite{Onogi:2005cz}%
\nocite{Ohta:2006qz}\cite{Damgaard:2007be}.}.
In ref. \cite{Takimi:2007nn}, 
Sugino's alternative lattice formulation 
\cite{Sugino:2003yb}--\nocite{Sugino:2004qd}%
\nocite{Sugino:2004uv}\cite{Sugino:2006uf}
was shown to follow 
from Catterall's by restricting the degrees of freedom of 
the complexified fields while preserving the supercharge. 
{}Furthermore, in ref. \cite{Damgaard:2007eh}, 
the formulations provided by the so-called link approach 
\cite{D'Adda:2004jb}--\nocite{D'Adda:2005zk}\cite{D'Adda:2007ax}
were also
shown to be equivalent to those of orbifolding.

Very recently, Catterall has shown that the orbifolded lattice gauge 
theories for two-dimensional $\cN=(2,2)$ SYM theory 
and four-dimensional $\cN=4$ SYM theory 
can be derived from topologically twisted continuum theories
using the geometrically discretization scheme without additional 
complexfication of fields \cite{Catterall:2007kn}. 
Together with the previous results \cite{Damgaard:2007xi}, 
this fact strongly suggests that Catterall's prescription 
for constructing a lattice theory with exact supersymmetry
from a continuum gauge theory is equivalent to  
that of orbifolding in general. 
The purpose of this note is to prove that it is indeed the case. 
In the following, we consider a general continuum gauge theory 
satisfying certain conditions and construct a lattice theory 
by means of orbifolding.  
In this way, we {\em derive} directly from orbifolding
a set of rules to construct 
the lattice action from a continuum action. 
We see that the rules we obtain are precisely those 
of the geometrical discretization scheme. The crucial $U(1)$
symmetries that generate the $d$-dimensional lattice in the
orbifolding formalism are behind the geometric picture which
emerges. In fact, as we will show, the rules are more general
and applicable to a theory with fields of more general tensor structure 
than that considered by Catterall. 

Let us start with a continuum gauge theory with gauge group $U(K)$ 
defined on the $d$-dimensional Euclidean flat space-time.
{}First, we impose certain conditions to the continuum action: \\
\noindent
\underline{\bf ASSUMPTIONS}
\begin{enumerate}
 \item The action is Lorentz invariant and consists of 
{\em complex} covariant derivatives $\cD_\mu$ and (bosonic and/or fermionic) 
tensor fields, $\{f^\pm_{\mu_1 \cdots \mu_p}\}$: 
\begin{align}
 S_{\rm cont.}
&=S_{\rm cont.}[\cD_\mu,\bcD_\mu,\{f^\pm_{\mu_1 \cdots \mu_p}\}] \nn \\
&\equiv \int d^dx \Tr \cL\left(
 \cD_\mu(x),\bcD_\mu(x),\{f^\pm_{\mu_1 \cdots \mu_p}(x)\}\right), 
\label{continuum action}
\end{align}
where $\cD_\mu(x)$ is associated with a complex (not hermitian) connection 
$\cA(x)$, 
$\bcD_\mu(x)$ is defined through the complex conjugate of $\cA_\mu(x)$,
$\overline{\cA}_\mu(x)=\cA^\dagger_\mu(x)$, 
and the trace is taken over the gauge indices. 
We further assume all the fields $\{f^\pm_{\mu_1\cdots \mu_p}(x)\}$
 are in the adjoint representation of $U(M)$.
\item The theory is assumed to have at least $U(1)^d$ symmetry and 
the complex covariant
derivative $\cD_\mu$ $(\bcD_\mu)$ possesses the $U(1)$ charges 
$\bfe_\mu$ $(-\bfe_\mu)$, where 
$\bfe_\mu$ is a set of $d$-dimensional linearly independent 
integer valued vectors.
\item We assume that the tensor field $f^\pm_{\mu_1\cdots\mu_p}$ 
has the $U(1)$ charge $\pm\sum_{i=1}^p\bfe_{\mu_p}$. 
Note that we can consider a more general field 
$f_{\mu_1\cdots\mu_p;\nu_1\cdots\nu_q}$ 
which has the ``mixed'' $U(1)$ charge 
$\sum_{i=1}^p \bfe_{\mu_i}-\sum_{j=1}^q\bfe_{\nu_j}$. 
This extension is straightforward but we only consider 
$f^\pm_{\mu_1\cdots\mu_p}$ for simplicity. 
\end{enumerate}

Under these assumptions, we dimensionally reduce the theory 
to zero dimension. 
At the same time, we enlarge the size of matrices from $K$ 
to $KN^d$ with a positive large integer $N$. 
As a result, all space-time dependence drops out and 
we obtain a matrix theory (a ``mother theory'') defined by the action
\begin{align}
 S_{\rm mother} 
&= S_{\rm mother}[\cA_\mu, \bar{\cA}_\mu, \{f^\pm_{\mu_1\cdots\mu_p}\}] \nn
 \\
&= \Tr \cL( 
i\cA_\mu, i\bar{\cA}_\mu, \{f^\pm_{\mu_1\cdots\mu_p}\}
), 
\label{mother action}
\end{align}
where $A_\mu$, $\bar{\cA}_\mu$ and $f^\pm_{\mu_1\cdots\mu_p}$ are 
complex matrices with the size $KN^d$. 
By assumption, the mother theory is invariant 
under the gauge transformation $\Phi \to g^{-1}\Phi g$,  
$(g\in U(KN^d))$ and the $U(1)$ transformation, 
$\Phi \to e^{iq_i \theta_i} \Phi$ 
$(0 \le \theta_i < 2\pi, \ i=1,\cdots,d)$,
where $\Phi\in\{\cA_\mu, \bar\cA_\mu, f^\pm_{\mu_1\cdots\mu_p}\}$ 
and $q_i$ $(i=1,\cdots,d)$ are the $U(1)$ charges of the field $\Phi$. 
In the orbifolding approach one starts with no {\em a priori} 
assumptions about $U(1)$ charge assignments. Different choices lead,
in general, to different lattice theories which can be classified
systematically \cite{Damgaard:2007be}. 
The action (\ref{mother action}) above corresponds
to the mother theory after one such choice.

This is exactly the situation where we can carry out 
the orbifolding procedure and produce a $d$-dimensional 
lattice action 
\cite{Cohen:2003xe}%
--\nocite{Cohen:2003qw}\cite{Kaplan:2005ta}. 
(See also ref. \cite{Damgaard:2007be}.)
Indeed, we can define an operator $P$ that 
projects out components that are not invariant 
under the $Z_N^d$ transformation. 
Here, for a matrix $f_{\bfmu;\bfnu}$ with $U(1)$ charge 
$\bfmu-\bfnu\equiv\sum_{i=1}^p\bfe_{\mu_i} -\sum_{j=1}^q \bfe_{\nu_j}$, 
we can parametrize the projected field as
\begin{align}
 P: \Phi_{\bfmu;\bfnu} \mapsto  P\Phi_{\bfmu;\bfnu} 
\equiv \sum_{\bfk \in \Z_N^d} {\Phi}_{\bfmu;\bfnu}(\bfk)\otimes
 E_{\fatk+\bfnu,\fatk+\bfmu}, 
\label{projected Phi}
\end{align}
where $\Phi_{\bfmu;\bfnu}(\bfk)$ is a complex matrix of size $K$, and 
we have defined 
\be
E_{\fatk,{\mathbf l}} ~=~ E_{k_1,l_1}\otimes \cdots \otimes
E_{k_d,l_d}. 
\qquad \Bigl( (E_{l,m})_{ij} \equiv \delta_{li}\delta_{mj} \Bigr)
\label{basis matrices}
\ee
The orbifold projection restricts fields in the mother
theory to those which are invariant under the operation of $P$. 
We obtain the orbifolded action 
by substituting (\ref{projected Phi}) into (\ref{mother action}):
\begin{align}
 S_{\rm orb} &= S_{\rm orb}[\cA_\mu(\bfk),\bar{\cA}_\mu(\bfk),
\{f^\pm_{\mu_1\cdots\mu_p}(\bfk)\}] \nn \\
&\equiv \Tr \cL(
 iP \cA_\mu, iP\bar\cA_\mu,\{P f^\pm_{\mu_1 \cdots \mu_p} \} ). 
\label{orbifolded action}
\end{align}
The lattice action is obtained by carrying out deconstruction 
\cite{Arkani-Hamed:2001ca} to the orbifold action 
(\ref{orbifolded action}), 
that is, by shifting the fields $\cA_\mu(\bfk)$ and 
$\bar\cA_\mu(\bfk)$ by $1/a$: 
\begin{align}
 \cA_\mu(\bfk) \to \frac{1}{a} + \cA_\mu(\bfk), \quad 
 \bar\cA_\mu(\bfk) \to \frac{1}{a} + \bar\cA_\mu(\bfk), 
\label{deconstruction}
\end{align}
where $a$ is interpreted as the lattice spacing.
Instead of this shift operation (\ref{deconstruction}), 
however, we here adopt a replacement of the fields as 
\cite{Unsal:2006qp} 
\begin{align}
 \cA_\mu(\bfk) \to \frac{1}{ia} e^{ia\cA_\mu(\bfk)} 
\equiv  -i\cU_\mu(\bfk), \nn \\
 \bar\cA_\mu(\bfk) \to \frac{1}{ia} e^{-ia\cA_\mu(\bfk)} 
\equiv  i\bcU_\mu(\bfk),
\label{replacement}
\end{align}
which is equivalent to (\ref{deconstruction}) 
to the leading order in the dimensionful quantity $a$,
$i.e.$ to the order of the naive continuum limit. 
As a result, we obtain a lattice action, 
\begin{align}
 S_{\rm lat} &= S_{\rm lat}[\cU_\mu(\bfk), \bcU_\mu(\bfk),
\{f^\pm_{\mu_1\cdots\mu_p}(\bfk)\}] \nn \\
 &\equiv \sum_{\bfk\in \Z_N^d} \Tr \cL_{\rm lat}(
 \cU_\mu(\bfk), \bcU_\mu(\bfk),
\{f^\pm_{\mu_1\cdots\mu_p}(\bfk)\} ) \nn \\
 &\equiv S_{\rm orb}[\cU_\mu(\bfk),-{\bcU}_\mu(\bfk),
\{f^\pm_{\mu_1\cdots\mu_p}(\bfk)\}], 
\label{lattice action}
\end{align}
where the trace in the second line is taken over a matrix 
with the size $K$.  
The naive continuum limit of this lattice theory 
is the gauge theory given by the action (\ref{continuum action}).

Let us now recall how the orbifolded matrix theory can be 
regarded as a lattice theory \cite{Cohen:2003xe}--\cite{Kaplan:2005ta}. 
Consider a matrix $\Phi$ of the size $KN^d$, 
which can be written as 
\begin{equation}
 \Phi=\sum_{\bfk,\bfl\in \Z_N^d} \Phi_{\bfk,\bfl} \otimes E_{\bfk,\bfl},
\end{equation}
where $\Phi_{\bfk,\bfl}$ is a matrix with the size $K$. 
The basic idea is that the $d$-vector $\bfk\in \Z_N^d$ 
labels a site of the lattice generated by the vectors
$\{\bfe_\mu\}_{\mu=1}^d$ as $\sum_\mu k_\mu \bfe_\mu$. 
Then the block $\Phi_{\bfk,\bfl}$ can be regarded as a variable 
living on an oriented link that goes from the site 
$\bfk$ to the site $\bfl$, which is expressed as $(\bfk,\bfl)$
in the following.
(The ``link'' $(\bfk,\bfk)$ corresponds to the site $\bfk$. )
Using this interpretation, it is easy to see that 
the lattice variables $\cU_\mu(\bfk)$, $\bcU_\mu(\bfk)$, 
$f_{\mu_1\cdots\mu_p}^+(\bfk)$ and $f_{\mu_1\cdots\mu_p}^-(\bfk)$ in 
(\ref{lattice action}) live on links $(\bfk,\bfk+\bfe_\mu)$, 
$(\bfk+\bfe_\mu,\bfk)$, $(\bfk,\bfk+\bfe_{\mu_1}+\cdots+\bfe_{\mu_p})$ 
and 
$(\bfk+\bfe_{\mu_1}+\cdots+\bfe_{\mu_p},\bfk)$, 
respectively.

As discussed in \cite{Cohen:2003xe}--\cite{Kaplan:2005ta}, 
the original gauge symmetry $U(KN^d)$ of the mother theory 
is broken to $U(K)^{N^d}$ 
by the orbifold projection (\ref{projected Phi}). 
More explicitly, the remaining 
gauge transformation is $\Phi\to g^{-1}\Phi g$ with
\begin{equation}
 g = \sum_{\bfk\in\Z_N^d} g(\bfk)\otimes E_{\bfk,\bfk}, 
\end{equation}
with $g(\bfk) \in U(K)$. 
Therefore, the lattice variables translate as 
\begin{align}
 \cU_\mu(\bfk) & \to g^{-1}(\bfk)\cU_{\mu}(\bfk)g(\bfk+\bfe_\mu), \nn \\
 \bcU_\mu(\bfk) & \to g^{-1}(\bfk+\bfe_\mu)\bcU_{\mu}(\bfk)
 g(\bfk), \nn \\
 f^+_{\mu_1\cdots\mu_p}(\bfk)&\to
 g^{-1}(\bfk)f^+_{\mu_1\cdots\mu_p}(\bfk)
g(\bfk+\bfe_{\mu_1}+\cdots+\bfe_{\mu_p}), \nn \\
 f^-_{\mu_1\cdots\mu_p}(\bfk)&\to
 g^{-1}(\bfk+\bfe_{\mu_1}+\cdots+\bfe_{\mu_p})
 f^-_{\mu_1\cdots\mu_p}(\bfk)
g(\bfk). 
\label{gauge trans}
\end{align}

Although the lattice action (\ref{lattice action}) is determined 
by substituting the decomposition (\ref{projected Phi}) into 
the mother action (\ref{mother action}), there is a short cut
to determine all terms of the lattice action. 
The key point is the $U(1)$ charges of the fields. 
{}For example, suppose that matrices $\Phi_\bfq$ and $\Psi_\bfr$ 
in the mother theory have $U(1)$ charges 
$\bfq\equiv\sum_{i=1}^q\bfe_{\mu_i}$ and 
$\bfr\equiv\sum_{j=1}^r \bfe_{\mu_j}$, respectively. 
As explained above, after the orbifold projection, 
the surviving blocks 
$\Phi_\bfq(\bfk)$ and $\Psi_\bfr(\bfk)$ 
can be interpreted as lattice variables living on links 
$(\bfk,\bfk+\bfq)$ and $(\bfk,\bfk+\bfr)$, respectively. 
On the other hand, the product $\Phi_\bfq\Psi_\bfr$ has the $U(1)$ charge 
$\bfq+\bfr$, so it is projected onto a (composite) variable 
living on the link $(\bfk,\bfk+\bfq+\bfr)$. 
Therefore, we can immediately see that this composite variable 
must be expressed as $\Phi_\bfq(\bfk)\Psi_\bfr(\bfk+\bfq)$ 
from the geometrical or the $U(1)$ charge point of view. 
An important application is the covariant derivative in the continuum 
theory (\ref{continuum action}). 
{}From the assumption of the continuum action, 
possible covariant derivatives appearing in the action 
are curl-like: 
\begin{align}
 \cD_\nu f^\pm_{\mu_1\cdots\mu_p}(x)
&=\del_\nu f^\pm_{\mu_1\cdots\mu_p}(x) 
 +i\left[\cA_\nu(x), f^\pm_{\mu_1\cdots\mu_p}(x)\right], \nn \\
 \bcD_\nu f^\pm_{\mu_1\cdots\mu_p}(x)&=
\del_\nu f^\pm_{\mu_1\cdots\mu_p}(x) 
 +i\left[\bar\cA_\nu(x), f^\pm_{\mu_1\cdots\mu_p}(x)\right], 
\label{curl}
\end{align}
or divergence-like:
\begin{align}
 \cD_{\mu_i} f^-_{\mu_1\cdots\mu_p}(x)&=
  \del_{\mu_i} f^-_{\mu_1\cdots\mu_p}(x)
   +i\left[\cA_{\mu_i}(x), f^-_{\mu_1\cdots\mu_p}(x) \right], \nn \\
 \bcD_{\mu_i} f^+_{\mu_1\cdots\mu_p}(x)&=
 \del_{\mu_i} f^+_{\mu_1\cdots\mu_p}(x)
   +i\left[\bar\cA_{\mu_i}(x), f^+_{\mu_1\cdots\mu_p}(x) \right]. 
\qquad (1\le i \le p)
\label{divergent}
\end{align}
Recalling that the charge assignment of the fields and 
the deconstruction (\ref{replacement}), we can show that 
the covariant derivatives (\ref{curl}) and (\ref{divergent})
in the continuum theory turn out to be 
\begin{align}
 \cD_\nu f^+_{\mu_1\cdots\mu_p}(x)&\to 
D_\mu^+ f^+_{\mu_1\cdots\mu_p}(\bfk)
\equiv 
\cU_\nu(\bfk) f^+_{\mu_1\cdots\mu_p}(\bfk+\bfe_\nu)
- f^+_{\mu_1\cdots\mu_p}(\bfk)\cU_\nu(\bfk+\bfmu), \nn \\
\cD_\nu f^-_{\mu_1\cdots\mu_p}(x)&\to 
D_\mu^+ f^-_{\mu_1\cdots\mu_p}(\bfk)
\equiv 
\cU_\nu(\bfk+\bfmu) f^-_{\mu_1\cdots\mu_p}(\bfk+\bfe_\nu)
- f^-_{\mu_1\cdots\mu_p}(\bfk)\cU_\nu(\bfk), \nn \\
 \bcD_\nu f^+_{\mu_1\cdots\mu_p}(x)&\to 
\bar{D}_\mu^+ f^+_{\mu_1\cdots\mu_p}(\bfk)
\equiv 
f^+_{\mu_1\cdots\mu_p}(\bfk+\bfe_\nu) \bcU_\nu(\bfk+\bfmu) 
- \bcU_\nu(\bfk)f^+_{\mu_1\cdots\mu_p}(\bfk), \nn \\
 \bcD_\nu f^-_{\mu_1\cdots\mu_p}(x)&\to 
\bar{D}_\mu^+ f^-_{\mu_1\cdots\mu_p}(\bfk)
\equiv 
f^-_{\mu_1\cdots\mu_p}(\bfk+\bfe_\nu) \bcU_\nu(\bfk) 
- \bcU_\nu(\bfk+\bfmu)f^-_{\mu_1\cdots\mu_p}(\bfk),
\label{forward diff}
\end{align}
and 
\begin{align}
\cD_{\mu_i} f^-_{\mu_1\cdots\mu_p}(x) &\to
D_{\mu_i}^- f^-_{\mu_1\cdots\mu_p}(\bfk) 
\equiv
\cU_{\mu_i}(\bfk+\bfmu-\bfe_{\mu_i})f^-_{\mu_1\cdots\mu_p}(\bfk)
-f^-_{\mu_1\cdots\mu_p}(\bfk-\bfe_{\mu_i})
\cU_{\mu_i}(\bfk-\bfe_{\mu_i}), \nn \\
\bcD_{\mu_i} f^+_{\mu_1\cdots\mu_p}(x) &\to
\bar{D}_{\mu_i}^- f^+_{\mu_1\cdots\mu_p}(\bfk) 
\equiv
f^+_{\mu_1\cdots\mu_p}(\bfk)\bcU_{\mu_i}(\bfk+\bfmu-\bfe_{\mu_i})
-\bcU_{\mu_i}(\bfk-\bfe_{\mu_i})f^+_{\mu_1\cdots\mu_p}(\bfk-\bfe_{\mu_i}), 
\label{backward diff}
\end{align}
respectively, 
where we have defined $\bfmu\equiv \sum_{i=1}^p \bfe_{\mu_i}$. 
We call the operators $D_\mu^+$ $(\bD_\mu^+)$ and $D_\mu^-$ 
$(\bD_\mu^-)$ the forward and backward covariant differences, respectively.

In summary, we have shown that
if a continuum gauge theory satisfies the stated assumptions, 
we can discretize it on a lattice generated by $\{\bfe_\mu\}$ 
by combining
dimensional reduction and the orbifolding procedure. 
Instead of carrying out explicit computation, 
we can be read off the lattice action (\ref{lattice action}) 
from the continuum action (\ref{continuum action})
by using the following prescription: \\
\noindent
\underline{\bf PRESCRIPTION}
\begin{enumerate}
\item The complex covariant derivatives $\cD_\mu$ and $\bcD_\mu$ 
become link variables $\cU_\mu(\bfk)$ and $\bcU_\mu(\bfk)$ 
on the links $(\bfk,\bfk+\bfe_\mu)$ and $(\bfk+\bfe_\mu,\bfk)$,
respectively, 
and the tensor fields $f^+_{\mu_1\cdots \mu_p}(x)$
and $f^-_{\mu_1\cdots \mu_p}(x)$
become lattice variables $f^\pm_{\mu_1\cdots \mu_p}(\bfk)$ 
living on the links $(\bfk, \bfk+\hmu_1+\cdots+\hmu_p)$ and 
$(\bfk+\hmu_1+\cdots+\hmu_p,\bfk)$, respectively.  
\item The gauge transformation of the lattice variables 
are given by (\ref{gauge trans}). 
\item  Curl-like complex covariant derivatives (\ref{curl}) 
become forward covariant differences (\ref{forward diff}). 
\item Divergence-like complex covariant derivatives 
(\ref{divergent}) become
backward covariant differences \!(\ref{backward diff}). 
\end{enumerate}
These are nothing but generalizations of the 
geometrical discretization rules proposed 
by Catterall \cite{Catterall:2004np}.
We have shown 
that they  follow directly
from orbifolding; both procedures always give the same lattice theory. 
We emphasize the novel point that the nature of lattice variables 
is uniquely determined not by the tensor structure 
{\em per se} but by the $U(1)$ charges of the fields. 
{}For example, let us consider the action 
of four-dimensional $\cN=4$ SYM theory in the form \cite{Catterall:2007kn}, 
\begin{align}
 S=\int d^4x \Tr \Bigl(&\left|\cD_\mu,\cD_\nu\right|^2
+\frac{1}{2}[\cD_\mu,\bcD_\mu]^2
+\frac{1}{2}[\phi,\bphi]^2
+\left(\cD_\mu \phi\right)\left(\bcD_\mu \bphi\right)
-\chi_{\mu\nu}\cD_{[\mu}\psi_{\nu]} \nn \\
&-\bpsi_\mu\cD_\mu \bareta -\bpsi_\mu[\phi,\psi_\mu]
-\eta[\bphi,\bareta]
-\frac{1}{2}\e_{\mu\nu\rho\sigma}\chi_{\rho\sigma}\bcD_\mu \bpsi_\nu
-\frac{1}{2}\e_{\mu\nu\rho\sigma}\chi_{\mu\nu}[\bphi, \chi_{\rho\sigma}]
\Bigr). 
\label{4D N=4}
\end{align}
If we assign $U(1)$ charge $\bfe_\mu$ $(-\bfe_\mu)$ to $\cD_\mu$ 
$(\bcD_\mu)$, we should assign $\bfe_\mu$ to $\psi_\mu$ 
by supersymmetry. Then the $U(1)$ charges for
the fields 
$\phi$, $\bphi$, $\eta$, $\bareta$ and $\bpsi_\mu$ 
are automatically determined to be 
$-\bfe_5$, $\bfe_5$, $0$, $-\bfe_5$ and $\bfe_5-\bfe_\mu$,
respectively, having defined $\bfe_5\equiv\bfe_1+\cdots\bfe_4$.
Therefore, the fields 
$\phi$, $\bphi$, $\bareta$ and $\bpsi_\mu$ should be 
written as $\phi_{\mu\nu\rho\sigma}$, $\bphi_{\mu\nu\rho\sigma}$, 
$\bareta_{\mu\nu\rho\sigma}$ and 
$\e_{\mu\nu\rho\sigma}\psi_{\nu\rho\sigma}$ 
in our notation, and the assignment on a lattice 
is uniquely determined to be the same as suggested in ref.
\cite{Catterall:2007kn}. 

We conclude this paper by making some comments. 
{}First, we call it a ``generalized'' 
geometrical discretization prescription because we do not restrict 
the tensor fields to be $p$-forms.
If the continuum theory contains only anti-symmetric 
tensor fields, the orbifolding procedure makes 
a $p$-form field $f_{\mu_1\cdots\mu_p}$ to be 
a lattice variable that
lives on a link $(\bfk,\bfk+\bfe_{\mu_1}+\cdots+\bfe_{\mu_p})$
or equivalently a $p$-cell $(\bfk;\bfe_{\mu_1},\cdots,\bfe_{\mu_p})$. 
In this case, the obtained lattice theory 
is ``local'' in the sense that all the variables live 
in a $d$-dimensional unit cell. 
This gives the original prescription of the geometrical 
discretization scheme.
However, we can apply the procedure described in this paper 
to a theory containing general tensor fields. 
The lattice theory so obtained might contain variables 
on links connecting non-nearest neighbor sites such as 
a double links, etc. So it is more general.

Second, we have not concentrated on exactly preserved
lattice supersymmetries in this paper. 
In fact, supersymmetry is irrelevant in the above argument, 
we have only used the assumption 
that fields carry the adjoint representation
of the gauge group. This is as expected from an earlier argument
due to Aratyn et al. \cite{Aratyn:1984bd}. 
However, when the continuum theory is supersymmetric, 
it is clear that the supercharges which have zero $U(1)$ charges 
are preserved by the orbifold projection \cite{Damgaard:2007eh}. 
Indeed, all supersymmetric lattice theories so far 
constructed by orbifolding 
and by the geometrical discretization scheme share this property. 

Third, when there are more than $d$ global $U(1)$ symmetries 
in the continuum theory, there is an ambiguity in the assignment 
of the $U(1)$ charges, and as a result, 
we can construct infinitely many different lattice theories 
whose formal continuum limit is the same. 
A typical example is two-dimensional $\cN=(4,4)$ SYM 
theory, whose action can be written as
\begin{align}
S = \frac{1}{g^2}\int d^2 x \Tr
\Bigl( 
&\left|[\cD_\mu,\cD_\nu]\right|^2
+ \frac{1}{2}[\cD_\mu,\bcD_\mu]^2
+(\cD_\mu\phi)(\bcD_\mu\bphi) + \frac{1}{2}[\phi,\bphi]^2 \nn \\
&+ \psi_\mu\bcD_\mu \eta
+\bpsi_\mu\cD_\mu \bareta
+\frac{1}{2} \xi_{\mu\nu}\cD_{[\mu}\psi_{\nu]}
+\frac{1}{2} \bxi_{\mu\nu}\bcD_{[\mu}\bpsi_{\nu]} 
\nn \\
&+\bareta[\bphi,\eta]+\bpsi_\mu[\phi,\psi_\mu]
+\frac{1}{2}\bxi_{\mu\nu}[\bphi,\xi_{\mu\nu}] 
\Bigr), 
\label{2D N=(4,4)}
\end{align}
where $\mu,\nu=1,2$, 
$\cD_\mu$ and $\bcD_\mu$ are complex covariant derivatives, 
$\phi$ and $\bphi$ are scalar fields and 
$\eta$, $\bareta$, $\psi_\mu$, $\bpsi_\mu$, $\xi_{\mu\nu}=-\xi_{\nu\mu}$
and $\bxi_{\mu\nu}=-\bxi_{\nu\mu}$ are fermionic fields. 
Apart from the manifest $U(1)^2$ symmetry with the charge assignment, 
\begin{center}
\begin{tabular}{c|cccccccccc}
 & $\cD_\mu$ & $\bcD_\mu$ & $\phi$ & $\bphi$ & $\eta$ & $\bareta$ & $\psi_\mu$ 
 & $\bpsi_\mu$ & $\xi_{12}$ & $\bxi_{12}$ \\
\hline
 $U(1)_1\times U(1)_2$ & $\fate_\mu$ & $-\bfe_\mu$ & $0$ & $0$ 
          & 0 & $0$ & $\bfe_\mu$ & $-\bfe_\mu$ 
          & $-\fate_1-\fate_2$ & $\fate_1+\bfe_2$ 
\end{tabular}\ , 
\end{center}
this theory has in addition two $U(1)$ symmetries, $U(1)_3\times U(1)_4$,  
whose charge assignments are given by 
\begin{center}
\begin{tabular}{c|cccccccccc}
 & $\cD_\mu$ & $\bcD_\mu$ & $\phi$ & $\bphi$ & $\eta$ & $\bareta$ & $\psi_\mu$ 
 & $\bpsi_\mu$ & $\xi_{12}$ & $\bxi_{12}$ \\
\hline
 $U(1)_3$  & 0 & 0 & 1 & -1  
          & 0 & 1 & 0 & -1 
          & 0 & 1 \\
 $U(1)_4$  & 0 & 0 & 0 & 0  
          & 1 & -1 & -1 & 1 
          & 1 & -1 
\end{tabular} .
\end{center}
Therefore, by adding the charges of $U(1)_3$ and $U(1)_4$ to 
those of $U(1)_1$ and $U(1)_2$, we can obtain infinitely many
charge assignments to the fields, and there are
correspondingly infinitely many lattice formulations%
\footnote{
One would then relabel the Lorentz indices of the fields 
corresponding to these different charge assignments.
}.
Note that we can obtain {\em supersymmetric} lattice theories 
by tuning the $U(1)$ charge of at least one of the fermionic 
field to be zero. 
The finite list of such theories are classified in \cite{Damgaard:2007be}.

{}Finally, as pointed out in the literature, 
the geometrical discretization scheme and, equivalently, 
the orbifolding procedure, naturally give rise
to Dirac-K\"ahler fermions on a lattice 
\cite{Catterall:2005eh}%
--\nocite{Kawamoto:1999zn}\nocite{Kato:2003ss}\cite{Kato:2005fj}. 
Indeed, 
Dirac-K\"ahler fermions can be defined on a lattice 
by using the correspondence 
between differential forms and co-chains 
\cite{Becher:1982ud}\cite{Rabin:1981qj}.
This correspondence gives a beautiful geometrical 
description of lattice fermions, and there is ample
evidence that they are very closely linked to exactly preserved
supersymmetries on the lattice. It remains to be shown 
explicitly why the orbifolding procedure  
always appears to give rise to such Dirac-K\"ahler fermions. 
Another outstanding
question to be answered concerns the addition of matter
multiplets to these theories. The geometrical rules seem to
lend themselves to matter carrying
other representations than just the adjoint. From the point
of view of orbifolding this is far from trivial
\cite{Endres:2006ic}. If, as we expect, there also here will
be an exact correspondence between the geometrical rules of
discretization and orbifolding this may give new insight into
orbifolded theories with matter in different representations.  

\vspace{0.5cm}
\noindent
{\sc Acknowledgement:}~ 
We thank S.~Hirano for useful discussions.
S.M. also acknowledges support from 
JSPS Postdoctoral Fellowship for Research Abroad.

\bibliographystyle{JHEP}
\bibliography{refs}

\end{document}